\documentclass[amsmath,amssymb,prl,twocolumn,showpacs]{revtex4-1}

\usepackage{amsmath}
\usepackage[applemac]{inputenc}			

\usepackage{graphicx}
\usepackage{amssymb}
\usepackage{bm}
\newcommand{\be}{\begin{equation}} 
\newcommand{\ee}{\end{equation}}
\usepackage{color}

\begin{document}

\title{Scaling Laws for the Slip Velocity in Dense Granular Flows}

\author{Riccardo Artoni}
\email{riccardo.artoni@unipd.it}
\author{Andrea C. Santomaso}
\author{Massimiliano Go'}
\author{Paolo Canu}
 
\affiliation{Dipartimento di Ingegneria Industriale, Universit\`{a} di Padova. Via Marzolo 9, 35131 Padova, Italy.
}%

\begin{abstract}
In this Letter, the 2-dimensional dense flow of polygonal particles on an incline with a flat frictional inferior boundary is analyzed by means of contact dynamics discrete element simulations, in order to develop boundary conditions for continuum models of dense granular flows. 
We show the evidence that the global slip phenomenon deviates significantly from simple sliding: a finite slip velocity is generally found for shear forces lower than the sliding threshold for particle-wall contacts.
We determined simple scaling laws for the dependence of the slip velocity on shear rate, normal and shear stresses, and material parameters.
The importance of a correct determination of the slip at the base of the incline, which is crucial for the calculation of flow rates, is discussed in relation to natural flows.
\end{abstract}
                              

\maketitle

\begin{figure}[!t]
\centering{
	\includegraphics[height=5 cm]{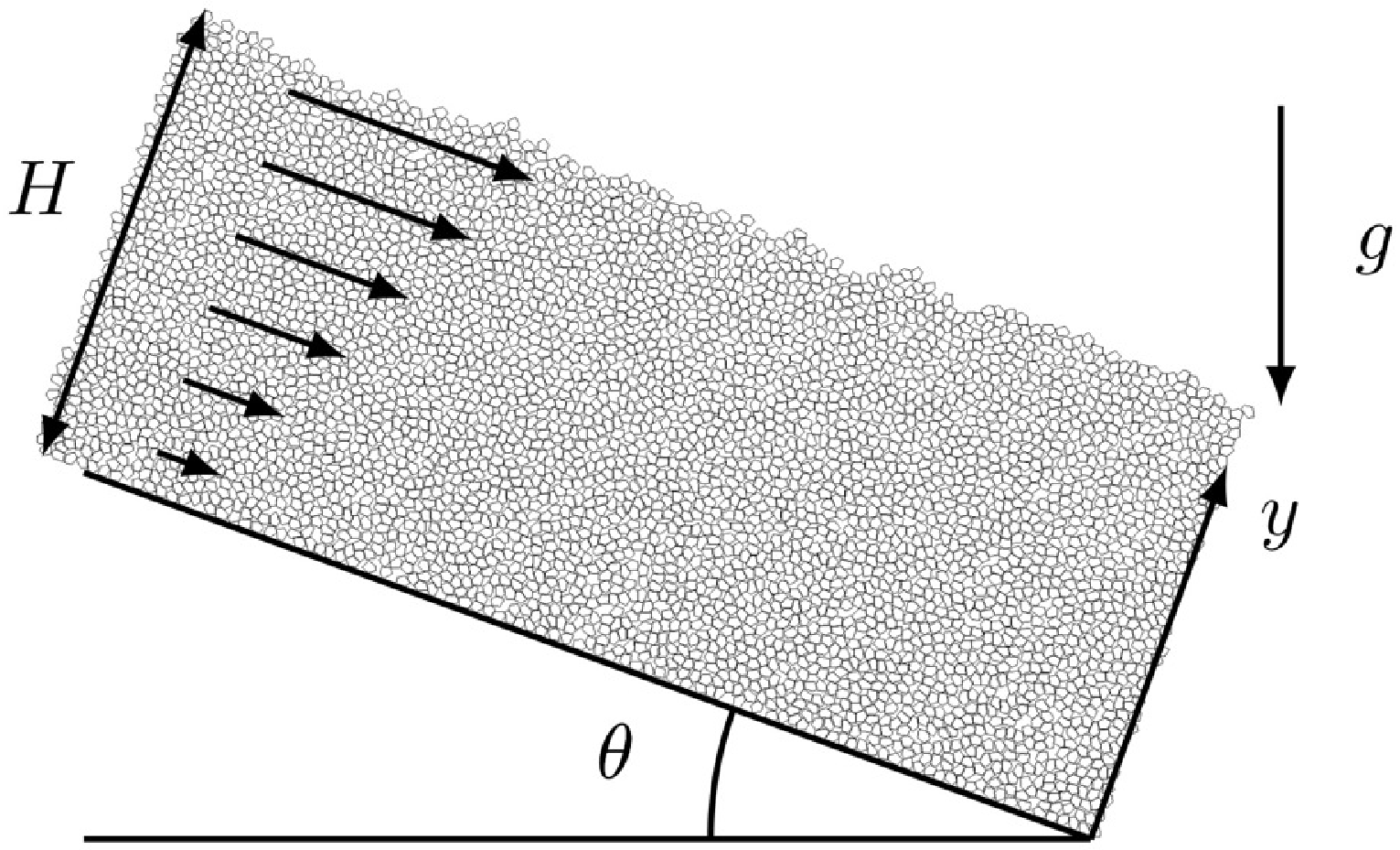}
	\includegraphics[height=5 cm]{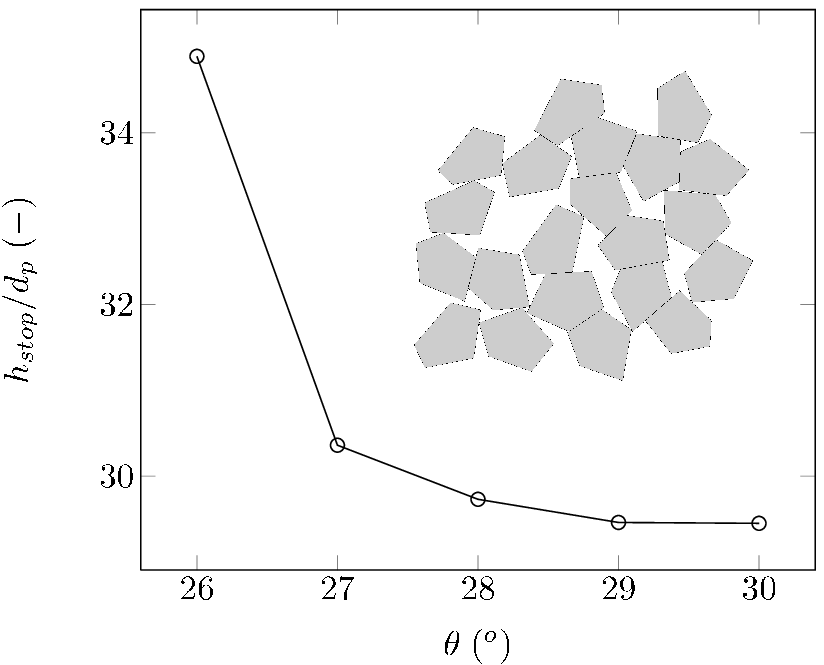}
}

	\caption{\label{fig1} (top) Sketch of the inclined chute geometry. (bottom)  $h_{stop}$ curve obtained with parameters $\rho_p=1000\; \mbox{kg/m}^3$,  $\mu_{pw}=0.8$, and sketch of an ensemble of particles to clarify the shape chosen (elongated ones). }
\end{figure}

Despite the large number of studies devoted to understanding the rheology of dense granular flows, the boundary behavior - {\em i.e.} the interaction of an ensemble of flowing particles with a wall -  is still poorly understood.  The issue is relevant for nearly all the situations in which granular materials are processed (silo or chute flows, sampling, die filling and compression) or are observed in nature (avalanches, landslides). A complete understanding of the boundary behavior requires us to study the relationships among the amount of wall slip, stresses and deformation rates in the material, possibly arriving at a quantitative average characterization to be used for characterization and for continuum modeling. This was extensively done with respect to rapid, dilute granular flows\cite{richman1,richman2, jenkins}, but the results from those analyses cannot be applied to dense, slow flows of granular materials since the phenomenology is largely different. Long-lasting contacts, dynamics of force chain networks, and dissipation mainly due to friction are the most significant distinctions between dense and collisional flow regimes.\\
From the practical point of view, in the dense flow of granular materials in silos and hoppers, wall friction is usually described by means of the effective wall friction coefficient $\mu_w=\frac{\sigma_{T}}{\sigma_{N}}$, which is a bulk, not a particle property, where $\sigma_{T}$ is the shear stress and $\sigma_{N}$ is the normal stress in the direction perpendicular to the wall. Such a coefficient, often presented as the angle of wall friction (defined as $\tan^{-1} \mu_w$), is not necessarily a constant property of the couple of particle and wall materials\cite{tuzun88,artoni09}. 
In presence of shear, in a recent work we argued \citep{artoni09} that shear-induced fluctuations of the force network could determine a dependence of the effective wall friction coefficient on flow properties, such as slip velocity, shear rate and stresses. The main point of the theory is that the stronger the phenomenon of force chains breaking and forming in the material, the larger will be the deviation from the condition of steady sliding  for the particles at a boundary, implying generally that $\mu_w<\mu_{pw}$, where $\mu_{pw}$ is the particle-wall friction coefficient, and suggesting the emergence of a dependence of the effective wall friction coefficient on the flow properties. In this Letter we prove the evidence of these flow dependent boundary conditions, and extend the scaling proposed in \citep{artoni09}.\\
In the following, the {\em dense} flow of granular materials composed of irregular particles down a  flat frictional inclined plane (as sketched in Fig. \ref{fig1}) is studied by means of discrete numerical simulations. 
The inclined chute configuration was chosen in this work due to the interesting feature that, if a continuum framework is valid, the effective wall friction coefficient can be in principle a known quantity. Indeed, under the hypothesis of steady state and neglecting inertial terms, the two dimensional momentum balance equations are $\partial_x \sigma_{xx}+ \partial_{y}\sigma_{xy}=\rho g \sin \theta$ and  $\partial_x \sigma_{yx}+ \partial_{y}\sigma_{yy}=-\rho g \cos \theta$. With the assumption of fully developed flow ($\partial_x =0$), and assuming that at the free surface $y=H$,  $\sigma_{ij}=0$, and that the density is constant along $y$, the $xy-$ and $yy-$ stress components in the geometry are given by $\sigma_{xy}=\rho g \sin \theta (y-H)$ and $\sigma_{yy}=- \rho g \cos \theta (y-H)$. In relation to the hypothesis of fully developed flow, the  $xx-$ and $yx-$ components are not known \emph{a priori}. Given that $\sigma_{yy}$ is the normal stress in the direction perpendicular to the wall, the effective wall friction coefficient is given by $\mu_w=\left| \sigma_{xy}/\sigma_{yy}\right|_{y=0}=\tan \theta$. Note that the effective bulk friction coefficient, which is defined as \cite{cruz05} $\mu^*=\tau/p=(\sigma_{xy}+\sigma_{yx})/(\sigma_{xx}+\sigma_{yy})$, will be equal to $\mu_w$ only if $\sigma_{xy}=\sigma_{yx}$ (which is a quite usual result of micromechanical investigations) and $\sigma_{yy}=\sigma_{xx}$ (which is not in fact obvious). In this case, if a simple sliding boundary condition was valid, with the macroscopic boundary condition coinciding with the microscopic one, this would imply for $\tan\theta<\mu_{pw}$ a no-slip behavior, for $\tan\theta=\mu_{pw}$ a steady sliding behavior, and for $\tan\theta>\mu_{pw}$  an unsteady accelerating motion at the wall. According to these results, the inclined chute is a benchmark configuration where effective wall slip behavior can be tested, since the inclination angle determines the value of the effective wall friction coefficient.

Discrete element simulations were carried out using the Contact Dynamics method (CD) developed by  J. J. Moreau and Michel \citet{jean99} and implemented in the open source platform LMGC90 \citep{renouf04}. The contact dynamics approach is used by a large community of researchers; its capability to reproduce the dynamics of granular materials was {studied} in a variety of situations spanning from flow in a rotating drum \citep{renouf05}, quasistatic shearing \citep{azema09}, granular column collapse \citep{staron05b}, {force transmission in packings \citep{azema07}, and flow down inclined chutes \citep{staron08,lois05b}.}
A detailed description of the {contact dynamics} approach can be found in the literature cited. To summarize the main aspects, the method considers perfectly rigid, nonoverlapping, circular (spherical) or polygonal (polyhedric) particles, which are free to translate and rotate. During particle contact, friction is accounted for by means of Coulomb's model: if the tangential force $T$ is lower than the Coulomb threshold $\mu_p N$, where $\mu_p$ is the microscopic, interparticle coefficient of friction, and $N$ is the normal force at the contact, the particle will not slip; otherwise, if $T\geq \mu_p N$, the two particles will slip with respect to each other. Collisions between particles are also accounted for by means of Newton's coefficient of restitution, which is defined as $e_p=u^+/u^-$, where $u$ is the particle velocity before ($-$) and after ($+$) contact (generally a normal and a tangential coefficient of restitution are defined, with $u$ being respectively the normal and tangential velocity). The same models for friction and collisions apply also to particle-wall contacts, which in principle have different values of the coefficients, say, $\mu_{pw}$ and $e_{pw}$. Within the approach, other interactions such as cohesive forces can also be introduced.\\ 
{In this work } a two-dimensional inclined chute was simulated, filled with up to 5000 particles. As the focus was on the behavior of irregular particles, polygonal particles were chosen. For spherical particles on a flat frictional inclined plane it is known that shear is localized in the first layer of particles due to rolling of the grains\cite{louge01b}. The behavior of irregularly shaped particles, where rotations are more difficult, and which can represent a wide variety of actual materials, is expected to be different and deserves attention. As \citet{azema07} discuss, using polygons instead of disks strongly influences the force transmission patterns through shape anisotropy and facetedness. In this study we considered regular pentagons deformed by means of an elliptic factor (see Fig. \ref{fig1} for examples  of particle shapes). This particular shape was chosen as a model of irregular particle with strong angularity. In order to investigate the effect of particle shape, some simulations were also carried out with regular pentagons. In order to prevent the formation of crystalline zones, the initial sample was slightly polydisperse, uniformly distributed in a range $d_p\pm 10 \% d_p$, where $d_p$ is the average particle diameter (defined as $\sqrt{4 A_p /\pi}$, where $A_p$ is the area of the irregular particle) corresponding in this work to $d_p=1 $ mm.
The description of the boundary behavior for more idealized systems such as nearly monodisperse disks or spheres flowing over ordered substrates can be more difficult since the phenomenology of these flows is rather rich, due to ordering or deordering patterns that are probably closely related to the peculiar geometrical nature of those particles \cite{silbert}. Also, systems with high polydispersity can be difficult to treat because of the emergence of segregation phenomena. For the sake of finding a master curve describing boundary behavior, many parameters were varied such as wall friction ($\mu_{pw}=$$0.63,0.65,$$0.67,0.7$$,0.8,0.9,1.0,$$1.03,$$1.05$$,1.07$$,1.08$),   flow depth ($H/d_p\sim 30, 40, 80, 100$), and inclination angle ($\theta= 26,27,28,29,30\; ^{o}$). In order to check the expected gravity and mass scalings, some simulations were also run with different values of gravity ($g=19.62,98.1\;m/s^2$) and particle density ($\rho_p= 5000,10000\; \mbox{kg/m}^3$, with $\rho_p=1000\; \mbox{kg/m}^3$ being the base case).
Interparticle and particle-wall interactions were chosen as completely inelastic ($e_p=e_{pw}=0$), with Coulomb friction ($\mu_p=0.3$) for long-lasting contacts. More than 40 simulations were performed. After an initial transient, when it was verified that the system was at steady state by analyzing instantaneous flow profiles and wall stresses, average values of velocity, stresses, and solid fraction were computed at several positions at the wall, as space-weighted time averages \cite{babic97}.
In order to avoid possible force imbalances coming from wrong treatment of the walls in the averaging procedure \cite{luding01f}, the strategy used in this work is to consider the force between a particle and a boundary as if it was shared by two particles, one of which is the real particle, the other is the same particle mirrored with respect to the boundary \citep{zhu02}. Regarding the flow regime realized in the simulations, the inertial number $I=\dot \gamma d_p /\sqrt{p/\rho_p}$ was verified in the range $5\times 10^{-3} - 1.5 \times 10^{-1}$ at the wall, corresponding to a dense flow. 
\begin{figure}[!t]
\centering{
	\includegraphics[width=\columnwidth]{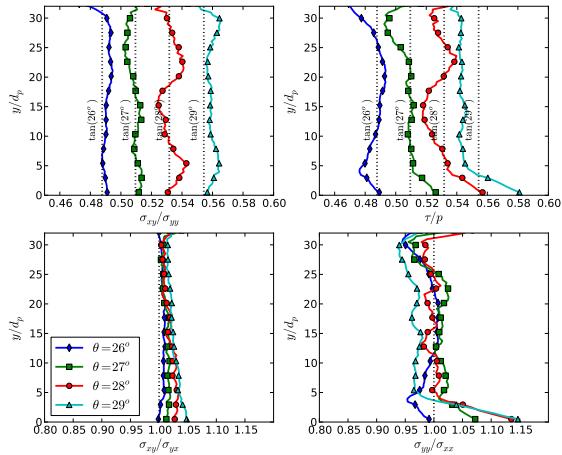}
}
\caption{\label{fig2a} Profiles of average stress ratios: (top left) $\sigma_{xy}/\sigma_{yy}$, (top right) $\mu^*=\tau/p$, (bottom left) $\sigma_{xy}/\sigma_{yx}$, (bottom right) $\sigma_{yy}/\sigma_{xx}$ , for  $\mu_{pw}=0.8$, $H/d_p \sim 30 $, $\rho_p=1000\; \mbox{kg/m}^3$,  for different values of the chute inclination $\theta$. }
\end{figure}
\begin{figure}[!t]
\centering{
	\includegraphics[width=\columnwidth]{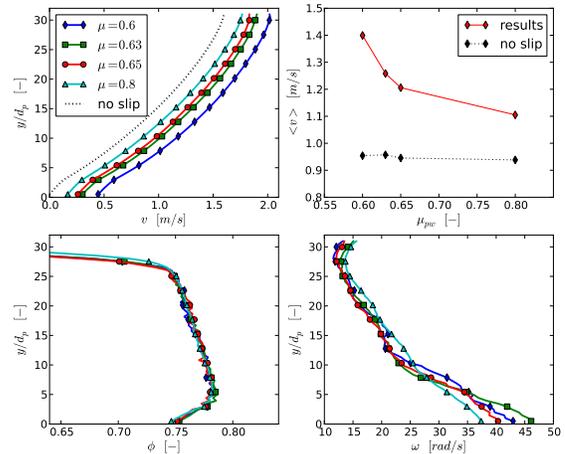}
}
\caption{\label{fig2} Profiles of average (top left) velocity,  (bottom left) solid fraction, (bottom right) particle rotation velocity, for  $\theta=29\; ^{o}$, $H/d_p \sim 30 $, $\rho_p=1000\; \mbox{kg/m}^3$, for different values of the particle-wall friction coefficient $\mu_{pw}$. (top right) Dependence of the average velocity on wall friction and comparison with hypothetical average velocity obtained assuming no-slip at the wall.}
\end{figure}\\
At first, in order to characterize the bulk behavior of the model material, the curve $h_{stop} (\theta)$, describing the depth of the flow below which the material jams, which depends on the inclination angle \cite{pouliquen99b}, was computed (it is reported in Fig. \ref{fig1}). When compared to previous results for disks (e.g. \cite{staron08}), the curve confirms that irregular particles have a higher resistance to flow; i.e., they tend to start jamming at higher values of the inclination angle.
Analyzing the results concerning stress profiles along $y$, an example of which is given in Fig. \ref{fig2a}, it is possible to conclude that the effective wall friction coefficient is approximately given by the tangent of the inclination angle, $\left|\sigma_{xy}/\sigma_{yy}\right|\approx \mbox{const.}\approx \tan \theta$, while the effective bulk friction coefficient $\mu^*$ is not always equal to $\tan \theta$. This is mainly due to the rather strong deviation from 1 of the normal stress ratio $\sigma_{yy}/\sigma_{xx}$.   On the other hand, the shear stress ratio $\sigma_{xy}/\sigma_{yx}$ deviates only slightly from 1, allowing us to consider $\sigma_{xy}=\sigma_{yx}$ an acceptable assumption. Regarding variations of stresses while varying $\mu_{pw}$ (not shown), stress profiles are not significantly affected by the friction coefficient of the wall, and satisfy the continuum momentum balance equations. This supports the validity of the continuum approach. 
Figure \ref{fig2} shows flow profiles for different values of the particle-wall friction coefficient. 
It appears that bulk profiles of shear and solid fraction are nearly unaffected by changes in the particle-wall friction coefficient. Shear rate has the typical $(H-y)^{3/2}$ behavior. Solid fraction is lower near the wall, then increases with $y$ in a 5 $d_p$ wide layer and decreases again up to the free surface. The rotational velocity is higher near the wall and decreases with $y$. Rolling influences the translational velocity of the grains at the wall:  slip at the wall is in principle due to two contributions, sliding and rolling, such that $v_{slip}=v_{slip,\; sliding }+ v_{slip,\;rolling}$, where the rolling contribution to the translational motion depends on the local average rotation $\omega$ through the expression $v_{slip,\;rolling}=\omega d_p /2$ (computing the velocity at the center of gravity of the first particle). In the simulations it was generally verified that $v_{slip,\; rolling }<0.1\; v_{slip,\;sliding}$, so that the rolling contribution to the translational motion is small with respect to the purely sliding contribution.
Contrary to what happens to bulk profiles, varying the particle wall friction coefficient does affect the slip velocity: the velocity profiles have the same shape, just shifted by a certain (and different for different values of $\mu_{pw}$) slip velocity at the base. This is a strong result since for values of $\mu_w\sim \tan \theta$ which are generally lower than $\mu_{pw}$, a certain slip is observed, which is not compatible with a simple sliding boundary condition as described previously in this Letter. A corollary of this result is that the total flow rate will be generally higher than the value obtained by assuming a classical sliding boundary condition, which for $\mu_w<\mu_{pw}$ would imply a zero sliding velocity. The difference between numerically predicted flow rate and the flow rate obtained assuming a simple sliding boundary condition is also shown in Fig. \ref{fig2}: the real flow rate (i.e. obtained from numerical experiments) can be 50 \% higher than expected from simple Coulomb sliding.  It appears that due to friction-induced shear, macroscopic behavior deviates from microscopic, contact-scale behavior: instead of $\mu_w=\mu_{pw}$, a nonconstant effective wall friction coefficient $\mu_w$ is found, implying that the continuum boundary condition is something intermediate between no-slip and steady-sliding friction and therefore cannot be described by simple models of sliding friction. These claims were the theoretical assumptions of our previous work \cite{artoni09}, and are now supported by evidence from numerical simulations accounting for detailed particle-wall mechanics.\\ 
Due to the evident deviation from simple sliding behavior, the dependence of the slip velocity on the system variables is not \emph{a priori} known. In \cite{artoni09}, based on a simple stochastic model, we suggested that an algebraic relation between dimensionless numbers $v_{slip}/\sqrt{p/\rho_p}$ and  ${\mu_w}/ {\mu_{pw}}$ could constitute a first simple model qualitatively describing the main finding, i.e. the existence of an intermediate behavior between no-slip and Coulomb friction. However, we verified that the scaling law proposed in \cite{artoni09} gives only an approximated description of the simulation results, most likely because the deformation of the flowing material was not considered in that simple model. In order to adapt the framework to a more complex situation it is better to improve the scaling $\sqrt{p/\rho_p}$ with $\dot \gamma d_p$, which was shown \cite{lois05b,lemaitre09} to be a universal scaling  for rigid grain systems,  independently of the value of the inertial number. Results confirm the need for this substitution, supporting the fact that $\sqrt{p/\rho_p}$ cannot be considered a quasistatic velocity scale.
This means that we look for a description of the amount of slip in the form of an algebraic relation
$\frac{v_{slip}}{\dot \gamma d_p}=f\left(\frac{\mu_w} {\mu_{pw}}\right) $. It is interesting to note that $\frac{v_{slip}}{\dot \gamma d_p}$ can be interpreted as a dimensionless slip length in the sense of  Navier.\\
As Figure \ref{fig4} shows, the results from the simulations shape a monotonic master curve which can be well described by the simple function $\frac{v_{slip}}{\dot \gamma d_p}= A \left(\frac{{\mu_w} /{\mu_{pw}}}{1-{\mu_w} /{\mu_{pw}}}\right)^{B} $, with best-fit parameters $A=2.2$, $B=0.42$ for elongated pentagons and  $A=4.23$, $B=0.33$ for regular pentagons. This is another important result: it seems that the nontrivial wall slip behavior can be described in the dense regime by algebraic relations between dimensionless numbers ${v_{slip}}/{\dot \gamma d_p}$ and  ${\mu_w}/ {\mu_{pw}}$, of which a tentative form was determined and proven to be succesful against contact dynamics data. 
By looking at Fig. \ref{fig4}, the deviation from simple sliding and the dependence of slip on flow and wall properties can be appreciated. At first, the curve is consistent in its limits with a simple sliding law: for example, in a stress-controlled experiment such as inclined chute flow, a very high particle-wall friction coefficient (i.e. $\mu_w/\mu_{pw}\rightarrow 0$) implies a no-slip behavior with finite shear ($v_{slip}/\dot \gamma d_p\rightarrow 0$). On the other hand $\mu_w/\mu_{pw}\rightarrow1$ implies a perfect sliding, with finite velocity and zero shear. But the behavior is not restricted to these two limits as in simple sliding: for $0<\mu_w/\mu_{pw}<1$ a finite  slip velocity and a finite shear are always found, with an increasing slip (and/or a decreasing shear) when increasing $\mu_w/\mu_{pw}$. The dependence on particle shape is then an expected result since it is reasonable that more rounded, less elongated particles will show a higher flowability, and thus a larger amount of slip at the wall. The fact that different types of particles follow the same behavior when studied through our framework is very encouraging for the development of realistic boundary conditions.
\begin{figure}[!t]
\centering{
	\includegraphics[height=5.8 cm]{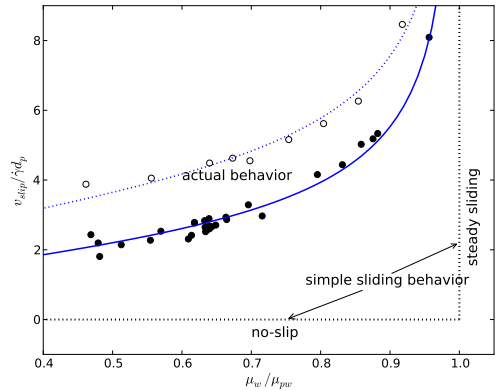}
	}
		\caption{\label{fig4} Dimensionless slip velocity $\frac{v_{slip}}{\dot \gamma d_p}$ vs rescaled effective wall friction coefficient $\frac{\mu_w} {\mu_{pw}}$. Symbols represent numerical data (filled circles correspond to elongated particles, open circles to regular ones), while lines are the corresponding fitting curves obtained with the functional form discussed in the text.}
\end{figure}
The proposed scaling law was determined for inclined chute simulations. The final scope of the research should be the determination of a universal scaling, i.e., a law that does not depend on the geometrical configuration, and should be valid also for real scale three-dimensional flows. For this reason a program of 3D simulations and dedicated experiments in various geometries is under development in our laboratory.. As a concluding remark, we want to stress that the choice of boundary conditions when modeling granular flows can be a very critical step: for example, here it was shown (see Fig. \ref{fig2}) that flow rates in inclined chute flows can be even 50 \% higher than expected from simple Coulomb sliding. The importance of this issue is obvious when thinking about natural flows down inclines such as, for example, rock avalanches, for which the determination of the runout distance is a very important issue. The understanding and modeling of appropriate continuum boundary conditions for dense granular flows as attempted in this Letter are therefore a major research scope not only for fundamental but also for practical reasons.
\acknowledgements
The authors wish to thank Fr\'{e}d\'{e}ric Dubois, Mathieu Renouf, and collaborators for interesting discussions and the help with LMGC90.

\end{document}